\documentclass[prb,showpacs,twocolumn,superscriptaddress,aps,a4paper]{revtex4}
\usepackage{pstricks,pst-node,pst-text,pst-3d,graphpap,pst-plot}
\usepackage{dcolumn}
\usepackage{amsmath}
\usepackage{graphicx}
\usepackage{latexsym}
\usepackage{amsfonts}
\usepackage{amssymb}
\DeclareGraphicsExtensions{.pdf,.gif,.jpg}

\newcommand{\be}{\begin{equation}}
\newcommand{\ee}{\end{equation}}
\newcommand{\beq}{\begin{eqnarray}}
\newcommand{\eeq}{\end{eqnarray}}

\begin{document}

\def\bbe{\mbox{\boldmath $e$}}
\def\bbf{\mbox{\boldmath $f$}}
\def\bg{\mbox{\boldmath $g$}}
\def\bh{\mbox{\boldmath $h$}}
\def\bj{\mbox{\boldmath $j$}}
\def\bq{\mbox{\boldmath $q$}}
\def\bp{\mbox{\boldmath $p$}}
\def\br{\mbox{\boldmath $r$}}
\def\bz{\mbox{\boldmath $z$}}

\def\bfzero{\mbox{\boldmath $0$}}
\def\bfone{\mbox{\boldmath $1$}}

\def\dr{{\rm d}}

\def\tb{\bar{t}}
\def\zb{\bar{z}}

\def\tgb{\bar{\tau}}

\def\bC{\mbox{\boldmath $C$}}
\def\bG{\mbox{\boldmath $G$}}
\def\bH{\mbox{\boldmath $H$}}
\def\bK{\mbox{\boldmath $K$}}
\def\bM{\mbox{\boldmath $M$}}
\def\bN{\mbox{\boldmath $N$}}
\def\bO{\mbox{\boldmath $O$}}
\def\bQ{\mbox{\boldmath $Q$}}
\def\bR{\mbox{\boldmath $R$}}
\def\bS{\mbox{\boldmath $S$}}
\def\bT{\mbox{\boldmath $T$}}
\def\bU{\mbox{\boldmath $U$}}
\def\bV{\mbox{\boldmath $V$}}
\def\bZ{\mbox{\boldmath $Z$}}

\def\bcalS{\mbox{\boldmath $\mathcal{S}$}}
\def\bcalG{\mbox{\boldmath $\mathcal{G}$}}
\def\bcalE{\mbox{\boldmath $\mathcal{E}$}}

\def\bgG{\mbox{\boldmath $\Gamma$}}
\def\bgL{\mbox{\boldmath $\Lambda$}}
\def\bgS{\mbox{\boldmath $\Sigma$}}

\def\bgr{\mbox{\boldmath $\rho$}}
\def\bgs{\mbox{\boldmath $\sigma$}}

\def\a{\alpha}
\def\b{\beta}
\def\g{\gamma}
\def\G{\Gamma}
\def\d{\delta}
\def\D{\Delta}
\def\e{\epsilon}
\def\ve{\varepsilon}
\def\z{\zeta}
\def\h{\eta}
\def\th{\theta}
\def\k{\kappa}
\def\l{\lambda}
\def\L{\Lambda}
\def\m{\mu}
\def\n{\nu}
\def\x{\xi}
\def\X{\Xi}
\def\p{\pi}
\def\P{\Pi}
\def\r{\rho}
\def\s{\sigma}
\def\S{\Sigma}
\def\t{\tau}
\def\f{\phi}
\def\vf{\varphi}
\def\F{\Phi}
\def\c{\chi}
\def\w{\omega}
\def\W{\Omega}
\def\Q{\Psi}
\def\q{\psi}

\def\ua{\uparrow}
\def\da{\downarrow}
\def\de{\partial}
\def\inf{\infty}
\def\ra{\rightarrow}
\def\bra{\langle}
\def\ket{\rangle}
\def\grad{\mbox{\boldmath $\nabla$}}
\def\Tr{{\rm Tr}}
\def\hc{{\rm h.c.}}

\title{Generalized waveguide approach to tight-binding wires:
Understanding large vortex currents in quantum rings}

\author{Gianluca Stefanucci}
\affiliation{Dipartimento di Fisica, Universit\`a di Roma Tor
Vergata, Via della Ricerca Scientifica 1, 00133 Rome, Italy}
\affiliation{Istituto Nazionale
di Fisica Nucleare, Laboratori Nazionali di Frascati, Via E. Fermi 40, 00044 Frascati, Italy}
\affiliation{European Theoretical Spectroscopy Facility (ETSF)}

\author{Enrico Perfetto}
\affiliation{Unit\`a CNISM, Universit\`a di Roma Tor Vergata, 
Via della Ricerca Scientifica 1, 00133 Rome, Italy}

\author{Stefano Bellucci}
\affiliation{Istituto Nazionale
di Fisica Nucleare, Laboratori Nazionali di Frascati, Via E. Fermi 40, 00044 Frascati, Italy}

\author{Michele Cini}
\affiliation{Dipartimento di Fisica, Universit\`a di Roma Tor
Vergata, Via della Ricerca Scientifica 1, 00133 Rome, Italy}
\affiliation{Istituto Nazionale
di Fisica Nucleare, Laboratori Nazionali di Frascati, Via E. Fermi 40, 00044 Frascati, Italy}


\begin{abstract}
We generalize the quantum waveguide approach to H\"uckel or 
tight-binding models relevant to unsaturated $\p$ molecular devices. 
A Landauer-like formula for the current density through {\em 
internal} bonds is also derived which allows for defining a local 
conductance. The approach is employed to study internal circular 
currents in two-terminal rings. We show how to predict the 
occurrence and the nature of large vortex currents in coincidence 
with vanishingly small currents in the leads. We also prove  
a remarkably simple formula for the onset of a vortex regime.
\end{abstract}


\maketitle


The Quantum Waveguide Approach (WGA) introduced in Refs. 
\onlinecite{x.1992,dj.1994} has been extensively used to study 
multiterminal mesoscopic structures, like quantum rings or quantum 
wires, possibly in the presence of impurities, magnetic fields, Rashba 
or Dresselhaus interactions. The basic idea of the WGA is to 
calculate the one-particle wavefunction 
$\vf_{n}(x)=a_{n}e^{ikx}+b_{n}e^{-ikx}$ in the $n$-th 
wire by imposing the continuity at each vertex, i.e., 
$\vf_{1}(0)=\vf_{2}(0)=\ldots=\vf_{{\cal N}}(0)$ (${\cal N}$  
being the number of intersecting wires at the vertex), and the additional 
condition $\sum_{n=1}^{{\cal N}}\vf'_{n}(0)=0$,  
which implies current conservation at the vertex (the derivative is 
taken along the incoming direction). It is worth noticing 
that current 
conservation is actually fulfilled by the weaker condition
\be
\sum_{n=1}^{{\cal N}}\vf'_{n}(0)=r\vf_{1}(0), 
\label{cc}
\ee 
where $r$ is an arbitrary real number. To our knowledge such 
arbitrariness has never been discussed.

In recent years it became possible to attach aromatic molecules or 
atomic chains to leads. These structures are geometrically similar to 
their mesoscopic counterparts and call for a H\"uckel-like or 
tight-binding (TB) generalization of the WGA due to the inadequacy of the 
continuum free-particle description. In the discrete case the 
difficulty stems from the fact that we cannot impose a condition on 
the derivative of the wavefunction. Approaches based on 
Green's functions\cite{mkr.1994}, iterative procedures\cite{ebgzr.2006} or 
source-and-sink-potentials\cite{gez.2007,pf.2008}
have been proposed but none of them is directly related to the WGA.

It is the purpose of this work to show how to generalize the WGA to 
the TB case (TBWGA) and to use the method to predict the occurrence of 
large vortex currents in quantum rings observed for the first time in Ref. 
\onlinecite{ebgzr.2006}. The continuum case is
recovered by a proper limit of the TB
parameters and allows us to understand the physical meaning
of the real constant $r$ in Eq. (\ref{cc}).

\begin{figure}[htbp]
\includegraphics*[width=.285\textwidth]{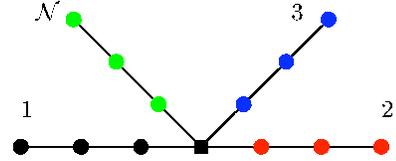}
\caption{Illustration of a vertex with ${\cal N}=4$.} 
\label{vertex}
\end{figure}

{\em Tight binding WGA}: We consider a generic system consisting of 
TB chains with 
at least one end-point in common. In Fig. \ref{vertex}  we
illustrate a vertex of the system with
${\cal N}=4$ intersecting chains. 
Let $\ve_{n}$ be the onsite energy of the $n$-th chain,
$t_{n}$ the hopping parameter
between nearest neighbor sites and $\ve_{V}$ the
onsite energy of the vertex.
We denote with 
$\q_{\ve n}(j)=a_{\ve n}e^{ik_{n}j}+b_{\ve n}e^{-ik_{n}j}$ the amplitude on 
the $j$-th site of the $n$-th chain of an eigenstate of energy
$\ve=\ve_{n}+2t_{n}\cos k_{n}$. As in the continuum case the continuity of the wave function yields ${\cal N}-1$
independent equations 
\be
\q_{\ve 1}(0)=\q_{\ve 2}(0)=\ldots=
\q_{\ve{\cal N}}(0).
\label{cont}
\ee
The additional condition, which plays the role 
of Eq. (\ref{cc}) in the WGA, is obtained by projecting on the vertex site
the stationary Schr\"odinger equation:
\be
\sum_{n=1}^{{\cal N}}t_{n}\q_{\ve n}(1)+\ve_{V}\q_{\ve 1}(0)=\ve\q_{\ve 1}(0).
\label{shdis}
\ee
Equations (\ref{cont}) and (\ref{shdis}) provide ${\cal N}$
independent equations for each vertex of the system.
Thus, for a system having ${\cal V}$ vertices with ${\cal N}_{i}$,
$i=1,\ldots,{\cal V}$, intersecting chains at the $i$-th vertex the 
above procedure yields $\sum_{i}{\cal N}_{i}$ equations.
Letting ${\cal P}$ be the
number of chains with both end-points belonging to the set of vertices and ${\cal Q}$
be the number of semiinfinite chains with one end-point
connected to a vertex, i.e., the number of terminals, we have 
$2{\cal P}+{\cal Q}=\sum_{i}{\cal N}_{i}$ and hence a degeneracy 
$D\leq {\cal Q}$ for each energy level (the number of unknown 
constants $\{a_{\ve n}\}$, $\{b_{\ve n}\}$ is $2{\cal P}+{\cal Q}+D$ where $D$ is the number of 
terminals for which $|\ve-\ve_{n}|<2|t_{n}|)$, 
as it should be.

To recover the WGA we employ a three-point discretization
of the kinetic term (our argument does not rely on this specific way
of discretizing). Then, $t_{n}=-1/(2\D^{2})$ and
$\ve_{n}=1/\D^{2}$, where $\D$ is the spacing between two points of
the continuum wire (we use atomic units). The amplitudes $\q_{\ve n}(1)$ correspond to the
amplitudes of the wavefunction at a distance $\D$ from the vertex. For
clarity we then rename $\q_{\ve n}(j)$ with $\q_{\ve n}(j\D)$ and
rewrite Eq. (\ref{shdis}) as
\be
\sum_{n=1}^{{\cal N}}\left[
\frac{\q_{\ve n}(\D)-\q_{\ve n}(0)}{\D}
\right]
-\D\tilde{\ve}_{V}\q_{\ve 1}(0)=
-2\ve\D\q_{\ve 1}(0),
\ee
where $\tilde{\ve}_{V}\equiv2\ve_{V}-{\cal N}/\D^{2}$. 
Taking the continuum limit $\D\ra 0$
we recover Eq. (\ref{cc}) provided that $\lim_{\D\ra 0}
\D\tilde{\ve}_{V}=r$, which implies $\tilde{\ve}_{V}\sim r/\D$. Thus, the constant 
$r$ is the amplitude of the $\d$-like potential $r\d(x)$ at the vertex
and is zero only for smooth potentials.

\begin{figure}[htbp]
\includegraphics*[width=.34\textwidth]{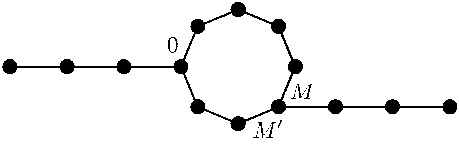}
\caption{Illustration of a ring device with $M=5$ and 
$N=8$. The last site of the lower arm is $M'=N-M$.} 
\label{ring}
\end{figure}

{\em Two terminal systems}:
We now focus on two terminal systems and obtain a {\em Landauer-like 
formula for the current density through a generic bond}. Let $H^{(0)}=
H_{\rm leads}+H_{\rm dev}+H_{\rm tun}$ be the Hamiltonian of the 
system in equilibrium.  The Hamiltonian of the left ($L$)
and right ($R$) leads is
\be
H_{\rm leads}=t\sum_{j<0}(
c^{\dag}_{j}c_{j-1}+\hc)+
t\sum_{j>0}(c^{\dag}_{j}c_{j+1}+\hc),
\ee
with nearest neighbor hopping $t$.
The device is described by $H_{\rm dev}=\sum_{nm=1}^{N} 
t_{nm}d^{\dag}_{n}d_{m}$ with real parameters $t_{nm}=t_{mn}$ and $N$ 
the total number of sites. The device is connected to the left
lead through site 1 and to the right lead through site $M \leq N$ 
(see, e.g., Fig. \ref{ring}).
The tunneling Hamiltonian is
\be
H_{\rm tun}=t_{L}(d^{\dag}_{1}c_{-1}
+\hc)+ t_{R}(d^{\dag}_{M}c_{1}+\hc).
\ee
We are interested in the long-time limit of the current density when 
an external bias $U_{\a}$ is imposed on lead $\a=L,R$.

At zero temperature the long-time limit of the density
matrix $\r_{x,x'}$, with $x,x'$ site indices of either the leads 
or the device, is given by the sum of a steady-state
contribution $\r^{(S)}_{x,x'}$ and a dynamical contribution.\cite{s.2007}
The steady-state contribution can be written in terms of left-going
eigenstates $|\q_{\ve R}\ket$ and right-going eigenstates $|\q_{\ve
L}\ket$  as
\be
\r^{(S)}_{x,x'}
=\sum_{\a=L,R}\int_{-2|t|+U_{\a}}^{\ve_{\rm F}+U_{\a}}
\frac{d\ve}{2\p}\q_{\ve\a}(x')
\q^{\ast}_{\ve\a}(x),
\label{rhoasy}
\ee
with $\ve_{\rm F}$ the equilibrium Fermi energy.
The states are normalized according to
$\bra\q_{\ve\a}|\q_{\ve'\b}\ket=2\p\d_{\a\b}\d(\ve-\ve')$.
Without loss of generality we choose $U_{L}>U_{R}$ and split Eq.
(\ref{rhoasy}) into three terms containing the contribution of the left-going evanescent
states with energy in the range $(-2|t|+U_{R},-2|t|+U_{L})$, the left-
and right-going current-carrying states with energy in the range
$(-2|t|+U_{L},\ve_{\rm F}+U_{R})$, and the right-going current-carrying states with
energy in the range $(\ve_{\rm F}+U_{R},\ve_{\rm F}+U_{L})$.
The evanescent states can be chosen 
real-valued since they are non-degenerate and the biased Hamiltonian
is invariant under time-reversal. Thus, the imaginary part of 
$\r^{(S)}_{x,x'}$ simplifies to
\beq
\Im[\r^{(S)}_{x,x'}]&=&
\int_{\ve_{\rm F}+U_{R}}^{\ve_{\rm F}+U_{L}}
\frac{d\ve}{2\p}
\Im[\q_{\ve L}(x')
\q^{\ast}_{\ve L}(x)]
\nonumber \\
&+&
\int_{-2|t|+U_{L}}^{\ve_{\rm F}+U_{R}}
\frac{d\ve}{2\p}\sum_{\a}\Im\left[\q_{\ve\a}(x')
\q^{\ast}_{\ve\a}(x)\right].
\label{imrho}
\eeq
Let us first consider the case in which both $x,x'$ are site indices of the same 
lead. The amplitude on the leads
$\q_{\ve \a}(j)\equiv \bra 0|c_{j}|\q_{\ve \a}\ket$ of a
normalized scattering state is
\beq
\q_{\ve L}(j)&=&\sqrt{\n_{L}(\ve)}\times \left\{
\begin{array}{ll}
    e^{iqj}+R_{\ve L}e^{-iqj} & j<0 \\
    T_{\ve L}e^{i\tilde{q}j} & j>0
\end{array}
\right.,
 \\
\q_{\ve R}(j)&=&\sqrt{\n_{R}(\ve)}\times \left\{
\begin{array}{ll}
    T_{\ve R}e^{-iqj} & j<0 \\
     e^{-i\tilde{q}j}+R_{\ve R}e^{i\tilde{q}j} & j>0
\end{array}
\right.,
\label{rsca}
\eeq
with $\ve=2t\cos(q)+U_{L}=2t\cos(\tilde{q})+U_{R}$, and
$\n_{\a}(\ve)=1/\sqrt{4t^{2}-(\ve-U_{\a})^{2}}$
the density of states in lead $\a$.
Exploiting current conservation and the orthogonality condition 
between left- and right-going  eigenstates, i.e.,
$R^{\ast}_{\ve L}T_{\ve R}/\n_{L}(\ve)+R_{\ve R}T^{\ast}_{\ve 
L}/\n_{R}(\ve)=0$,  it is straightforward to show that 
for $x=j>0$ and $x'=j+1$ the second term in Eq. (\ref{imrho}) 
vanishes while the first term reduces to the well known Landauer 
formula 
\be
\Im[\r^{(S)}_{j,j+1}]=\frac{1}{2|t|}
\int_{\ve_{\rm F}+U_{R}}^{\ve_{\rm F}+U_{L}}
\frac{d\ve}{2\p}|T_{\ve L}|^{2}\frac{\n_{L}(\ve)}{\n_{R}(\ve)}.
\label{land}
\ee
Below we show that the 
possibility of expressing $\Im[\r^{(S)}_{x,x'}]$ as an integral over 
the bias window is valid {\em for all sites} 
including those in the central device. Let us
express $\q_{\ve R}$ as a linear combination of $\q_{\ve\a}$ and 
the time-reversal state $\q^{T}_{\ve\a}=\q^{\ast}_{\ve\a}$
\be
\q_{\ve R}(x)=\sqrt{\frac{\n_{R}(\ve)}{\n_{L}(\ve)}}
\frac{1}{T^{\ast}_{\ve L}}\left[\q^{\ast}_{\ve L}(x)-R^{\ast}_{\ve L}
\q_{\ve L}(x)\right].
\label{psipsistar}
\ee
Extracting the transmission and reflection coefficients $T_{\ve R}$
and $R_{\ve R}$ one can easily verify that $\q_{\ve R}(x)$ is 
orthogonal to $\q_{\ve L}(x)$. Inserting Eq. (\ref{psipsistar}) into 
Eq. (\ref{imrho}) and exploiting current conservation for the 
right-going scattering state, one realizes that the imaginary 
part of $\sum_{\a}\q_{\ve\a}(x')\q^{\ast}_{\ve\a}(x)$ is identically 
zero and hence only states in the bias window contribute to $\Im[\r^{(S)}_{x,x'}]$. 
Thus, the long-time limit of
the current density $J_{nm}\equiv 2t_{nm}\Im [\r^{(S)}_{n,m}]$ 
through an
internal bond of the device connecting site $n$ to site $m$ can 
be expressed in a Landauer-like formula. 
In linear response $J_{nm}=G_{nm}(U_{L}-U_{R})$ and exploiting the 
above result the {\em local conductance} $G_{nm}$ is given by
\be
G_{nm}\equiv \frac{J_{nm}}{U_{L}-U_{R}}=
\frac{t_{nm}}{\p}\Im[\q_{\ve_{\rm F} L}(m)
\q^{\ast}_{\ve_{\rm F} L}(n)].
\label{bcnd}
\ee
We next specialize the analysis to devices consisting
of a TB ring and address the existence of vortex regimes.

{\em Conductances in two terminal rings}: We consider a ring with $N$ sites, see Fig. 
\ref{ring}. For notational
convenience we denote with $d_{\Uparrow n}$, $n=0,\ldots, M$ the
fermionic operators in the upper arm,
$d_{\Downarrow n}$, $n=0,\ldots,N-M$ the fermionic operators in the
lower arm, and identify $d_{\Uparrow 0}\equiv d_{\Downarrow 0}$, 
$d_{\Uparrow M}\equiv d_{\Downarrow N-M}$.
In terms of the operators $d_{\Uparrow n},\;d_{\Downarrow n}$ the
device Hamiltonian reads
\be
H_{\rm dev}=t\sum_{n=0}^{M-1} (d^{\dag}_{\Uparrow n}d_{\Uparrow n+1}+\hc)
+t\sum_{n=0}^{M'-1} (d^{\dag}_{\Downarrow n}d_{\Downarrow n+1}+\hc),
\ee
where $M'=N-M$ and the hopping $t$ is the same as in the leads. For simplicity we 
also set $t_{L}=t_{R}=t$ in the tunneling Hamiltonian.
We employ the TBWGA to calculate, e.g., the right-going eigenstates
$\q_{\ve L}$. Let
$\q_{\ve L}(j)=\sqrt{\n_{L}(\ve)}(e^{ikj}+R_{\ve L}e^{-ikj})$ be the amplitude on lead
$L$ ($j\leq0$)
and $\q_{\ve L}(j)=\sqrt{\n_{L}(\ve)}T_{\ve L}e^{ikj}$ be the 
amplitude on lead $R$ ($j\geq0$).
Similarly, the wavefunction on the upper arm of the ring has the form
$\q_{\ve L}(n)=\sqrt{\n_{L}(\ve)}(A_{\ve L}e^{ikn}+B_{\ve L}e^{-ikn})$
with $0\leq n\leq M$ while on the lower arm 
$\q_{\ve L}(n)=\sqrt{\n_{L}(\ve)}(C_{\ve L}e^{ikn}+D_{\ve L}e^{-ikn})$
with $0\leq n \leq N-M$.
According to the TBWGA the coefficients of $\q_{\ve L}$ are solution 
of 
\begin{widetext}
\be
\left(
\begin{array}{llllll}
    1 & 1 & 0 & 0 & -1 & 0 \\
    0 & 0 & 1 & 1 & -1 & 0 \\
    e^{ikM} & e^{-ikM} & 0 & 0 & 0 & -1 \\
    0 & 0 & e^{ik(N-M)} & e^{-ik(N-M)} & 0 & -1 \\
    e^{ik} & e^{-ik} & e^{ik} & e^{-ik} & -e^{-ik} & 0 \\
    e^{ik(M-1)} & e^{-ik(M-1)} & e^{ik(N-M-1)} & e^{-ik(N-M-1)} & 0 & -e^{-ik}
\end{array}
\right)
\left(
\begin{array}{l}
    A_{\ve L} \\ B_{\ve L} \\ C_{\ve L} \\ D_{\ve L} \\ R_{\ve L} \\ T_{\ve L}
\end{array}
\right)=
\left(
\begin{array}{l}
    1 \\ 1 \\ 0 \\ 0 \\ e^{ik} \\ 0
\end{array}
\right).
\label{linsys}
\ee
\end{widetext}

Left-going states $\q_{\ve R}$ can be computed in a similar manner and
it is straightforward to show that $T_{\ve L}=T_{\ve R}\equiv T_{\ve}$, which
is the transmittance of the system. The conductance through a bond is
given by Eq. (\ref{bcnd}). For any bond in the leads the conductance
is simply $G_{\ve}=|T_{\ve}|^{2}g_{0}$ where $g_{0}=1/(2\p)$ is the
quantum of conductance for spinless electrons. The
conductance on the upper and lower arm of the ring are
$G_{\Uparrow \ve}=(|A_{\ve L}|^{2}-|B_{\ve L}|^{2})g_{0}$ and
$G_{\Downarrow \ve}=(|C_{\ve L}|^{2}-|D_{\ve L}|^{2})g_{0}$.
From the above system of equations we obtain the following analytical
solution for the transmittance and the local conductance
\be
T_{\ve}=
\frac{ie^{ik(2M-N)}\sin^{2}(k) \sin(k\frac{N}{2})\cos(k(\frac{N}{2}-M))}
{\W(k)},
\label{trans}
\ee
\be
\frac{G_{\Uparrow \ve}}{g_{0}}=
\frac{\sin^{4}(k) \sin(k\frac{N}{2})\cos(k(\frac{N}{2}-M))\sin(k(N-M))
}{2|\W(k)|^{2}},
\label{condup}
\ee
\be
\frac{G_{\Downarrow \ve}}{g_{0}}=
\frac{\sin^{4}(k) \sin(k\frac{N}{2})\cos(k(\frac{N}{2}-M))\sin(kM)
}{2|\W(k)|^{2}},
\label{conddown}
\ee
with
\beq
\W(k)&=&
\frac{e^{-2ik(N-M+2)}}{16}
\left[
1-4e^{2ik}(1-e^{2ik})-e^{2ik(M+2)}\right.
\nonumber \\
 &+& \left.e^{2ikN}-2e^{ikN}
\left(e^{2ik}-1\right)^{2}
-e^{2ik(N-M+2)}
\right].
\eeq
It is easy to verify that current conservation is fulfilled since
$G_{\Uparrow \ve}+G_{\Downarrow \ve}=|T_{\ve}|^{2}g_{0}=G_{\ve}$. It
is also worth emphasizing that $G_{\ve}$ is bounded between $0$
and $g_{0}$ while $G_{\Uparrow \ve}$ and $G_{\Downarrow \ve}$ can be much
larger than $g_{0}$ and either positive or negative. We say that we 
are in a {\em vortex regime} if ${\rm sign}[G_{\Uparrow \ve}]=-{\rm sign}[G_{\Downarrow \ve}]$.
From Eqs. (\ref{condup},\ref{conddown}) we conclude that {\em for arms of different length a vortex regime
always exists}
since
\be
\frac{G_{\Uparrow \ve}}{G_{\Downarrow \ve}}=\frac{\sin(k(N-M))}{\sin(kM)}.
\label{vr}
\ee
Equation (\ref{vr}) is simple and transparent. The onset of
a vortex occurs for those values of the incident momentum
corresponding to an eigenenergy of either the isolated lower arm
$k^{({\cal L})}_{n}=\p n/(N-M)$ or of the isolated upper arm
$k^{({\cal U})}_{m}=\p m/M$.
In the following we characterize the vortex regime and show how to 
predict the occurrence of large ring currents in coincidence with a
vanishingly small current in the leads.

\begin{figure}[htbp]
\includegraphics*[width=.47\textwidth]{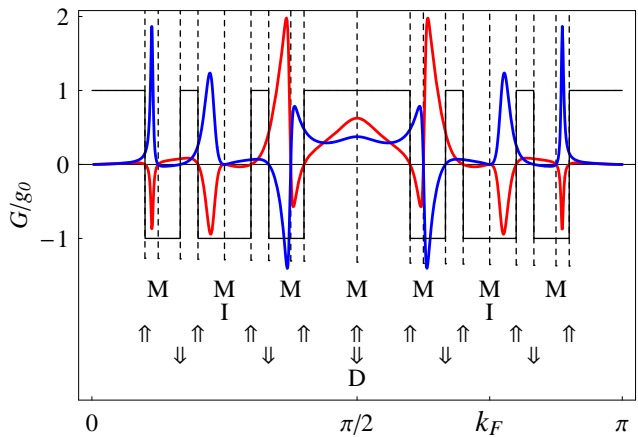}
\caption{(Color online) Local conductance $G_{\Uparrow \ve}/g_{0}$ (red), 
$G_{\Downarrow \ve}/g_{0}$ (blue) and $S_{\ve}\equiv 
{\rm sign}[\sin(k(N-M))/\sin(kM)]$ (black), see Eq. (\ref{vr}), 
versus the incident momentum $k$. The function $S_{\ve}$ is 1 in the 
laminar regime and -1 in the vortex regime. Vertical dashed lines are 
drawn in correspondence of $k=k^{{\cal M}}$ (M),  $k=k^{{\cal I}}$ 
(I), $k=k^{{\cal L}}$ ($\Uparrow$), $k=k^{{\cal U}}$ ($\Downarrow$), 
and of the zeros of $\W(k)$ (D).} 
\label{zeridoppi}
\end{figure}

{\em Vortex regime}: 
As already pointed out in Refs. \onlinecite{ycp.2003,pf.2008} the
transmittance $T_{\ve}$ has
two different kinds of zeros. The numerator in Eq. (\ref{trans}) vanishes either for energies
that exactly match an eigenvalue of the isolated ring, i.e.,
for $k^{({\cal M})}_{n}=2n\p/N$ ({\em matching momenta}) or for energies at which there is perfect destructive
interference at the right interface, i.e., for $k^{({\cal I})}_{m}=(2m+1)\p/(N-2M)$
({\em interference momenta}).
At these points the transmittance vanishes provided the denominator
$\W(k)\neq 0$. From the linear system in Eq. (\ref{linsys}) one can
easily show that $\W(k)$ cannot be zero at
the interference momenta since the wavefunction at the right
interface vanishes and hence
$T_{\ve}$ vanishes as well. As the numerator of $T_{\ve}$ goes to zero
as $\sim (k-k^{({\cal I})}_{n})$ the denominator $\W(k)$ has to be finite.
On the other hand, the denominator can vanish at the matching
momenta for special values of $M/N$. Expanding $\W(k)$ around
$k^{({\cal M})}_{n}$ one finds to first order
$\W(k)\approx -(e^{4iMn\p/N}-1)^{2}/16
+\g(n)(k-k^{({\cal M})}_{n})$,
with $|\g(n)|>\sqrt{N(2N-3)+M(4N+2)}>0$. Thus $\W(k)$ approaches zero
as $(k-k^{({\cal M})}_{n})$ for integer $2Mn/N$. In these cases the
simple zero of the denominator cancels the simple zero of the
numerator and $T_{\ve}$ is finite (unless $k^{({\cal M})}_{n}$ is an
interference momentum as well). Our condition for the cancellation of
zeros include the symmetric case $M=N/2$ already discussed in Ref.
\onlinecite{ycp.2003} as well as other cases, 
see Fig. \ref{zeridoppi} where $N=16$ and $M=6$ and the zero at 
$\p/2$ is cancelled by the denominator (the conductance $G_{\ve}$ has the 
same zeros as the transmittance since $G_{\ve}/g_{0}=|T_{\ve}|^{2}$).

From Eqs. (\ref{condup},\ref{conddown}) one can see that a zero of $G_{\ve}$
implies a zero of both $G_{\Uparrow \ve}$ and $G_{\Downarrow \ve}$.
Novel zeros, however, exist for the local conductances. 
Specifically, the numerator of $G_{\Uparrow \ve}$
vanishes for $k=k^{({\cal L})}_{n}$ while the numerator of $G_{\Downarrow \ve}$ vanishes for
$k=k^{({\cal U})}_{m}$, see Fig. \ref{zeridoppi}. Thus, the conductance $G_{\Uparrow \ve}$ is zero at
$k=k^{({\cal L})}_{n}$ unless $k^{({\cal L})}_{n}$ is also a zero of
the denominator. In this latter case $G_{\Uparrow \ve}$ remains
finite since it exists an integer $m$ such that $k^{({\cal L})}_{n}=
k^{({\cal M})}_{m}$ and hence $\sin(Nk/2)\sin(k(N-M))\sim
(k-k^{({\cal M})}_{m})^{2}$ which cancels the double zero of
$|\W(k)|^{2}$.\cite{note} In Fig. 
\ref{zeridoppi} this cancellation takes place at $k=\p/2$.
A similar reasoning apply to  $G_{\Downarrow \ve}$.
To summarize, the local conductances $G_{\Uparrow \ve}$, 
$G_{\Downarrow \ve}$ as well as the transmittance $T_{\ve}$ can have 
simple or double zeros while $G_{\ve}$ double or quadruple
zeros.

In accordance with the above analysis the onset of the vortex phase
occurs for incident momenta $k=k^{({\cal U})}\neq k^{({\cal
L})}$ or $k=k^{({\cal L})}\neq k^{({\cal U})}$. Below we show that in
the vortex regime there are special values of the incident momentum
yielding a circular ring current much larger than the current in the
leads. In order to quantify this effect we define the vortex function
\be
V(\ve)=\frac{G_{\Uparrow \ve}-G_{\Downarrow \ve}}{G_{\ve}}.
\ee
The modulus of $V(\ve)$ is in the range (0,1) when $G_{\Uparrow \ve}$ and
$G_{\Downarrow \ve}$ have the same sign (laminar regime) and is
greater than one in the vortex regime.
At the onset either $G_{\Uparrow \ve}$ or $G_{\Downarrow \ve}$
vanishes and $|V(\ve)|=1$. The vortex function diverges at the zeros of
the total conductance and the magnitude of the vortex can be
classified according to the nature of zeros of $G_{\ve}$. For double
zeros of $G_{\ve}$ (single zeros of $T_{\ve}$)
the difference $G_{\Uparrow \ve}-G_{\Downarrow \ve}$
has a single zero since a double zero would imply $k^{({\cal
U})}=k^{({\cal L})}$ which in turn implies $\W(k)=0$.
Expanding the conductances around the single zero $\ve_{s}$
of $T_{\ve}$ we find $V(\ve)\sim 1/(\ve-\ve_{s})$.
The current flowing in the ring changes direction as $\ve$ crosses $\ve_{s}$.
It is worth noting that in the 
neighborhood of $\ve_{s}$ the derivative $|G'_{\Uparrow/\Downarrow \ve}|$  can 
be very large (see Fig. 3 at $k_{F}=3\p/8$) and
$|G_{\Uparrow/\Downarrow \ve}| \sim g_{0} >> G_{\ve} \sim 10^{-1}g_{0}$.
Even more striking is the behavior of $V(\ve)$ around a double zero
$\ve_{d}$ of $T_{\ve}$. In this case $G_{\Uparrow \ve}-G_{\Downarrow 
\ve}$ also has a double zero\cite{note} and hence
$V(\ve)\sim 1/(\ve-\ve_{d})^{2}$. In this case the vortex does 
not change sign as $\ve$ crosses $\ve_{d}$.
Both divergences are rather remarkable since one could naively expect 
that $|G_{\Uparrow/\Downarrow \ve}|< G_{\ve}$ in accordance with  
Kirchoff's current laws of classical electromagnetism. 

We also explored the vortex regime for nonzero 
onsite energies $\ve_{\rm dev}$ on the ring and different couplings $t_{L/R}$.
Varying $t_{L/R}$ alters the shape of $G_{\Uparrow/\Downarrow}$
but preserves both the position and the nature of the zeros.
On the contrary, the position of the zeros changes as $\ve_{\rm dev}$ is varied 
but a vortex regime still exists.

In conclusion we have generalized the WGA to TB models and
obtained a Landauer-like formula for the current density through a 
generic bond. The TBWGA requires the {\em same} computational effort as the 
WGA. Employing the TBWGA in combination with the obtained 
expression of the local conductance we showed how to
predict the occurrence of large vortex currents. 
The existence of a vortex regime is rather robust and has to be attributed to 
the nontrivial topology of the system rather than to a 
specific choice of the parameters.

Acknowledgment: S.B. acknowledges partial support by the grant PRIN 2006.



\begin{thebibliography}{10}
\bibitem{note}
The conclusion that the local conductance in the rings remains finite
relies on the fact that it is not possible to have a triple zero in
the numerator of Eqs. (\ref{condup},\ref{conddown}), i.e., it is not
possible to verify $k^{({\cal M})}=k^{({\cal I})}=k^{({\cal U}/{\cal L})}$.
\end{thebibliography}
\end{document}